# Saudi Arabian Parents' Perception of Online Marital Matchmaking Technologies

ADEL Al-DAWOOD, University of Minnesota, USA, Al Imam Mohammad Ibn Saud Islamic University, Saudi Arabia

SERENE ALHAJHUSSEIN, Independent Researcher, Saudi Arabia

SVETLANA YAROSH, University of Minnesota, USA

Finding a date or a spouse online is usually considered an individualistic endeavor in Western cultures. This presents a challenge for collectivist non-Western cultures such as Saudi Arabia where choosing a spouse is viewed as a union of two families with parents of both spouses being heavily involved. Our work aims to investigate how Saudi Arabian parents view the utilization of technology by their young adults to seek potential spouses online. We report our findings of interviews conducted with 16 Saudi Arabian parents (8 fathers, 6 mothers and 1 couple). We generate qualitative themes that provide insights about how parents wanted to preserve their values, integrate technology into the traditional process and protect their young adults from potential harms. These themes lead to implications for designing suitable marital matchmaking technologies in Saudi Arabia and opportunities for future work.[1]

## 1 INTRODUCTION

Technology has been utilized to find dates or relationships for almost two decades [29], but not much work has explored the use of technology with other stakeholders being involved (i.e. family). While it is common that people seek dates and relationships individually, recent work has explored how it is different for "Eastern" populations where involving the family in the process makes it safer [65]. In addition, another study has explored the role of community in online dating platforms [49]. These studies have suggested that a collective experience on dating or matrimonial platforms might be more effective than an individualized experience. We encourage the Human-Computer Interaction (HCI) community to investigate how this could make the design of matchmaking technologies more inclusive as our world population is becoming more diverse.

As Saudi Arabia marriages usually follow a traditional process where both families are involved in the process, it is important to understand how this would translate to a matrimonial matchmaking system. A previous study [4] has shown that Saudi Arabians are interested in utilizing technology for marriage, but are unsure how to navigate it while preserving their cultural and religious values and involving their parents in the process. Other studies have explored how Saudi Arabians utilized matrimonial websites to seek a potential spouse [18] and how it might conflict with the traditional process [19]. For example, women in Saudi Arabia require their legal guardian's approval when deciding who they want to marry, which can vary based on many familial and social factors [55]. In addition, family members who are against the marriage can create obstacles for potential spouses[19]. Lastly, Saudi Arabians are obligated to please their parents and respect them based on Islamic teachings [2,4,37]. To the best of our knowledge, no previous work has explored Saudi Arabian parents' perspective on marital matchmaking technologies. For those reasons, our work aims to answer the following research questions:

**RQ1.** What concerns and views do Saudi Arabian parents have about marital matchmaking technologies?

**RQ2.** What is an effective role for Saudi Arabian parents to play in marital matchmaking technologies?

We investigated these questions through the analysis of 16 semi-structured interviews that we conducted with Saudi Arabian parents with the aim to contribute to the ongoing discussion in HCI on inclusive design and cross-cultural computing. In doing so, we aim to inform the design process for both creating new and appropriating existing technologies to support matchmaking in Saudi Arabia.

## 2 BACKGROUND

### 2.1 Traditional Marriage Process

To understand the context of our work, it is important to explain the traditional marriage process in Saudi Arabia. Refer to previous works [4–6,12,16] for more thorough explanations. The process can be summarized in the following 4 steps:

1. **Proposal:** The mother of the prospective groom approaches the mother of prospective bride to confirm that the families are suitable for each other and reach an initial agreement. Sisters and aunts are possible substitutes for mothers if needed.
2. **Social engagement:** This is similar to "dating" in the western context, but with the family aware of the couple's interactions and having relative control over them. For example, they can talk over the phone but in person meetings are chaperoned.
3. **Contract:** After the couple and families decide they are suitable for each other; a contract is usually drafted by a public notary, which makes the couple legally married.
4. **Wedding:** After the contract is drafted, the couple and their families start to arrange for the wedding party. It is a social announcement of the marriage and when the couple starts to live together.

These steps can vary depending on what parents find acceptable. For example, some parent believe direct interactions with potential spouses is limited or non-existent before the contract is drafted or the wedding night is over. On the other hand, other parents might allow direct interactions to start during or before the proposal. Breaking the contract before the wedding is considered as a divorce, so a relationship is considered to be "serious" once a contract is in place. A divorce enacted by breaking a contract affects future marriage prospects, however it may still be viewed as more favorable than a divorce after the wedding. Divorces after a wedding may be even more problematic for the woman due to a premium on women's virginity in conservative cultures. Divorces are generally stigmatized in Saudi Arabia and avoided, similar to other countries [25,42,50].

This work focuses on the potential role of technology during the first two steps, the proposal and social engagement. Parental involvement and the exchange of information are the main factors in these steps that are related to our research questions.

## 2.2 Stakeholders

The marriage process involves a number of stakeholders[2]:

1. **Parents:** They have the most influence over the process. Issues may arise if parents of the bride and groom have conflicting opinions. Parents might insist that they choose who their adult children marry or allow their adult children to marry whoever they wish. In some cases, the parents of the bride and groom may negotiate conditions about the marriage without consulting the bride and groom, such as the dowry and wedding arrangements.
2. **Groom:** He usually chooses who to marry based on his parent's recommendations. He could also suggest certain women for his parents to consider. Even though he may sever ties with his parents if he goes against their desires, he does not need their approval to marry legally.
3. **Bride:**  She usually waits for the groom's family to propose to her family and has the option to accept or reject the proposal. However, women in Saudi Arabia need the approval of their guardian (i.e. a direct male relative, e.g. a father) to marry legally [55]. This means that a guardians disapproval of a groom may hinder the marriage process. At the same time, the bride may view the approval of her guardian as protection, i.e. part of the vetting process.
4. **Public notaries:** It is common for Saudi Arabians to draft the contract with a public notary rather than in a courtroom. They usually serve as a mediator between the families and the courts and judges.
5. **Courts and Judges:** They usually handle the documentation and legal processes of the marriage. They may intervene when violations occur but are otherwise not involved. For example, a court may transfer the approval of the father to a judge if a case is made that the father is preventing his daughter(s) from marriage, also known as *adhl*.

## 2.3 Contemporary Gender Norms, Roles and Expectations in Saudi Arabia

In the last 2 years, Saudi Arabia has made major law reforms affecting gender equality. Saudi women were granted the right to drive on June 24th, 2018 [80]. In 2019, guardianship laws were changed to allow women to travel, marry, divorce, be a guardian for minor children, and represent themselves in court without the cooperation of a male guardian [81]. Gender segregation laws were also overturned in late 2019, allowing women and men to interact in public settings [79].

Even though Saudi Arabian women have gained more rights, they need to adhere to cultural and religious values that have been ingrained within Saudi Arabia. For example, registering a marriage or a divorce requires proving that a male guardian is abusing their authority in the case of *'adhl* or paying the dowry back in the case of *khul'*[3]. In addition, following through with *'adhl* or *khul* comes at the cost of possible family ties being severed as a result. In other cases, extended family members could intervene to annul a marriage based on tribal incompatibility. This decision is usually made to avoid violence that may occur against the potential spouses or between their families if the marriage is not annulled.

---

[2] Saudi Arabia follows Sharia law and marriage can only be between a man and a woman.
[3] Khul' is where a wife demands to be separated from her husband in Islamic Jurisprudence.

These changes in Saudi Arabian laws have affected the legal aspects of marriage and meeting their future spouse, however many cultural norms regarding matrimony remain unchanged. Cultural and religious values such as *bir alwalidayn* [2] play a vital role in the marriage process as most Saudi Arabians have personal, cultural, and religious reasons to maintain a healthy relationship with their parents when choosing who they marry. It is possible that young adults may avoid challenging the status quo publicly while privately being against it.

## 3 RELATED WORK

### 3.1 Cross-Cultural Online Matchmaking technologies and Privacy

Online dating has been around for almost 3 decades [61] and U.S. marriages that started online between 2005 and 2012 were associated with better marital satisfaction and were less likely to end up with divorce or separation compared to alternatives [24,34]. This topic is complicated to study but also needs to be explored more by the HCI community overall [22,23,29,75,78], but most previous work was done with western populations in mind (North America and Europe). We highlight previous work on online dating in general and relate it to work done on understudied populations.

Authenticity and trust are a constant concern when using technology and especially in the case of online matchmaking technologies. Lee and Bruckman found that users managed their identity differently between social media and online dating, but their social media identity was more authentic and "natural" [44]. The ability to view a friend list on social media was a method to measure the credibility of a user. Trust and credibility are a major concern with online dating especially when deception and self-exaggeration is more likely to occur in online dating platforms [32,67] and current mechanisms have not been successful with establishing trust [57]. Even though deception is expected in online dating, Hancock et. al. found that that its magnitude is usually small [35]. Nonetheless, this creates a struggle when evaluating potential romantic partners on an online dating platform as found by Zytko et. al [76].. They also found in another study that online dating coaches focused on moving the interaction offline as soon as possible. They believed that online evaluation of profiles was not effective at measuring a user's personality and meeting in person would be more effective [74,77]. On the contrary, Saudi Arabians found online settings to provide more information since their interaction with the opposite gender offline is limited [14,17]. Even though recent changes have made it easier to interact with the opposite gender in public settings, exchanging information for marriage publicly is usually not acceptable. It is preferred if it is done discretely, and online platforms are well suited for that. Nonetheless, online matchmaking platforms lack a community that would provide feedback and advice for users, which would lead them to seek third party community forums (i.e. Reddit) as found by Masden and Edwards [49].

In line with that, Nayak et. al. found that social matching based on previous relationships in online dating platforms increased the success rate of recommendations [56], but this depends on how success is measured in online dating platforms [48].  The definition of success is important for our population since it usually translates to marriage and not going on a first date or being engaged [16,48]. Another indicator of success in online dating was more self-disclosure as found by Handel and Shklovski [36], which was found to be associated with anonymity by Ma et. al. [46]. At the same time, there is the concern that anonymity would question the authenticity of self-disclosed information [45]. On the other hand, self-disclosing sensitive information such as a disability [60] or being HIV positive [36] might seem unfavorable for the user, but is also considered as a filtering mechanism to avoid negative outcomes. In addition, Ma et. al. found that online daters considered overlapping location data to be useful, even though it presented security and privacy concerns [47]. We can see that information sharing is useful, but only when it is trustworthy and does not violate users' privacy.

Preserving privacy is connected to preserving reputation and honor in Muslim and Arab cultures [7,27]. Privacy invasion online can be a serious threat to relationships in Arab societies, which is more prevalent for those with a mid to low socioeconomic background that might hold onto traditions more compared to those from a high socioeconomic background [62]. In Arab culture there is a perception that not sharing information with family and friends is a sign of mistrust, but this is also where they become vulnerable to privacy invasion threats. Abokhodair et al. defined privacy in Islam in terms of three aspects: *haq al-khososyah*, *hurma*, and *awrah* with each encompassing each other in the same order [3]. The first, *haq al-khososyah*, includes the right for an individual to protect their private life to preserve their reputation. The second, *hurma*, is meant specifically to protect the body from those who are not allowed to see it. The third, *awrah*, is more specific to intimate body parts. Saudi Arabians view their privacy as a right that is respected collectively in Muslim and Arab cultures, but may find it harder to achieve the same level of collective recognition of this value in the Western world [1,3]. Since most current social media systems are developed with Western values in mind, this creates a fundamental tension in the use of social media and makes controlling their privacy troublesome.

While Saudi Arabians are avid users of social media (e.g., 75% penetration rate on Twitter [82]), they may hesitate to reveal any personal information online. Previous research reveals that Saudi Arabian users are likely to use unidentifiable nicknames online and have multiple accounts to maintain more control over individual privacy [2,33]. These concerns also have gendered components. Social media is frequently viewed as a dangerous space for women [8] due to concerns over potential for blackmail if personal photos or information is gained using social media [3,10]. This usually means that women are less likely to use their personal photos on social media [9,21]. While using nicknames and having no personal photos may seem to undermine their agency in a western context, it allows them to use social media more freely in the Muslim and Arab context with no fear of repercussions. Use of social media and online dating increases more for women compared to men when they move to a western country where this stigma may be less salient [8,66]. Given these concerns and stigmas, Saudi women are only likely to use matrimonial websites after exhausting all possible traditional methods [4,15,18]. A previous study by Bajnaid et. al. found that Saudi Arabian men have a higher success rate of moving from online marital matchmaking technologies to offline interactions compared to Saudi Arabian women [18]. Personal agency is a concern for both genders when using technology for marriage in Saudi Arabia [18]. Nonetheless, Technology seems to provide more agency for women since they have less control over the traditional marriage process compared to men (Section 2.2).

The potential stigma of technologically-mediated matchmaking is evident in the fact that most Saudi Arabians report not knowing any acquaintances who have used matrimonial websites [15,18]. As this stigma is more prominent for women due to traditional courtship roles [4,15], men are more likely to be open about using matrimonial websites compared to women and were also more active on them [15,66]. Both Saudi men and women users of such sites agreed that they used them to find more choices compared to traditional means [4,15,19], but wanted to conform to cultural and religious norms while making use of such sites [10,11,33]. Even though marital matchmaking technologies are viewed as challenging social and religious norms, Saudi Arabian users find that their religious and cultural values shape how they view themselves online [14,17].

A study conducted by Al-Dawood et al. found that most Saudi Arabians showed interest in utilizing matchmaking technologies but were concerned about their parents accepting the use of matchmaking technologies and how to find a suitable way to collaborate with parents in finding and vetting potential partners that does not undermine young adults' agency [4]. One aspect of parental vetting of potential partners for marriage in Saudi Arabia is lineage [16]. For Saudi Arabians, this usually means differentiating between those who belong to a tribe and those who do not with some tribes preferring to marry within. A study done by Ortega and Hergovich [58] shows that interracial marriages increased in the US with the increase of online dating usage. In other words, marital matchmaking technologies could challenge the status quo of only marrying based on lineage compatibility that Saudi Arabian parents support [16]. This is a source of tension between parents and young adults since parents prefer monocultural marriages and young adults are willing to explore intercultural marriages while being aware that they might struggle [4]. As Saudi Arabian culture is becoming more technology-oriented, it might be possible that intercultural marriages would increase.

While previous investigations covered privacy and trust concerns while using online matchmaking technologies, the role of Muslim and Arab parents when marriage is conducted online is underexplored. What information do they consider appropriate or important to exchange when evaluating a potential partner? What kinds of cross gender interactions would they find acceptable? How does privacy and trust play a role in the information exchange and interaction between potential partners? In this work, we contribute a qualitative investigation seeking to address these questions.

### 3.2   Parent-Child Technological Interactions

Previous work has highlighted interactions between parents and their children when using technology [59,64,70,73]. While we understand young adults have different interactions with their parents compared to children, the difference is blurred between Muslim and Arab young adults and their parents given the religious and cultural expectations. We highlight previous work on parent-child technological relationships and explain the uniqueness of the Muslim and Arab context for these relationships.

Previous work has explored how parents and adolescents manage their use of technology. Technology is meant to enable teenagers to learn about making safe decisions on their own as restricting or monitoring their usage by their parents is not effective [13,30,71]. In other words, technology should be designed for them to be able to explore its usage properly and learn from their experiences with parents providing them with guidance. In one study, adolescents were better at making moral judgements with authoritative parents which had high expectations regarding appropriate use of technology but did not restrict usage [71]. In other studies, it was found that parental control actually increased online risks while granting teens more autonomy helped them learn from their negative experiences [30,31] and that developing resilience toward online risks is more effective than using restrictive measures to limit exposure to online risks [72]. In a study where parents monitored their children's activities, children were less likely to voluntarily disclose information, which had a negative effect on trust between children and their parents [41]. Teens prefer that their parents are not involved in their online usage to preserve their privacy. At the same time, teenagers are willing to give access to their information when it is an emergency. Teens are also more likely willing to share their information with parents if they were notified about it. In other cases, technology might provide a medium for brokering between parents and their youth that would improve communication channels between them [68]. In two studies done by Muñoz et al, they worked on exploring how technology can play a role in improving parent-adult children relationships by incorporating dialogue that would allow them to renegotiate their positions after children have left home and manage expectations for both ends [52,53]. For Arabs and Muslims, this may be slightly different as the renegotiation of positions and expectations is guided by religion and culture.

Previous work has highlighted the importance of family values in Islamic and Arab cultures. The concept of *bir alwalidayn* is ingrained in Islam and Arab culture and it requires obeying and respecting parents or other family elders into adulthood [2,4,37]. Most decisions are traditionally influenced by the parents or other family elders rather than by young adults [28,69]. Even when young adults make the decision on their own, they usually consult the parents about it [43]. This may introduce tensions with personal agency for young adults who want to make their own decisions but also respect the opinion of their parents, which may seem to be contradictory [4]. This tension is usually intensified when using technology that was designed with the individualistic and independent values of the West in mind. Muslim young adults may view respecting their parents' opinions as part of their personal agency as they choose to partake in it. In other cases, they may feel socially pressured to do so. Finding a balance between personal agency and respecting parents in Muslim and Arab cultures is a constant struggle online.

Some people may resolve this tension by keeping parts of their life separated. Some Muslims and Arabs present different aspects of self offline versus online to avoid conflict with their parents [3]. For example, they may hide information or photos that may reveal them at mixed-gender gatherings [7] in order to follow the social expectations set by their families [3,10]. When people employ this strategy, technology may increase the danger of context collapse, such as being tagged in content that would violate cultural or religious norms [3]. This may cause people to be extra vigilant that their posts or online traces cannot be used to harm their own or their family's reputation and honor [2,8,33]. For example, some users protect their family's reputation and honor by hiding their identities in their online activities [2,7,33]. One of the key implications of this strategy is that people may hesitate to use online tools for purposes that traditionally involve family involvement, particularly matchmaking. While these practices protect users' privacy and gives them more personal agency online, it alienates their family in the process.

However, while previous work has articulated the importance of family involvement, the specifics of this involvement and how it relates to other sociotechnical considerations of a potential matchmaking site remain unclear. Our work contributes an empirical investigation on how parents perceive online matchmaking technologies and what role they would play in them and investigating what types of interactions and information would be suitable or important in the online marital matchmaking process.

## 4 METHODS

### 4.1 Participants

Participants were initially recruited through an online sign up form that was distributed through multiple social media platforms. The main platform was Twitter, which was utilized by soliciting retweets from Saudi Arabian public figures to reach more participants. Another recruitment method was the first author personally reaching out to acquaintances that might be interested or know someone who was interested. We attempted to recruit participants from diverse backgrounds but were not successful. The regions our participants belonged to were limited to Riyadh (8), Makkah (4), Najran (2), Albaha (1) and Tabuk (1). It is worth noting that we Saudi Arabia has 13 regions, which means less than half the regions are represented in our sample. Most of our participants were highly educated with 10 of them having a masters or a PhD. We have an even distribution for income with middle class being the majority and lower middle class and upper middle class as outliers. Participants were as young as 38 years old and as old as 63 years old. For the background column, Qabali is having tribal linage while Arabi is not having tribal linage, which we will refer to in other parts of the paper for simplicity. In addition, a Hadhri lived a modern lifestyle (e.g. City) while a Badawi lived a simpler lifestyle (e.g. Rural). Table 1 below provides more info about our participants.

Table 1. Study participants Characteristics

| # | Region | Gender | Background | Education | Income | Age |
|---|--------|--------|------------|-----------|--------|-----|
| P1 | Riyadh | Male | Qabali Hadhri | Bachelors | Mid | 53 |
| P2 | Riyadh | Female | Qabali Hadhri | PhD | Mid | 46 |
| P3 | Riyadh | Male | Arabi Hadhri | Masters | Mid | 43 |
| P4 | Makkah | Female | Qabali Hadhri | Bachelors | Mid | 38 |
| P5 | Riyadh | Male | Arabi Hadhri | Highschool | Low | 63 |
| P6 | Riyadh | Male | Qabali Hadhri | Bachelors | Mid | 52 |
| P7 | AlBaha | Male | Qabali Badawi | Masters | Low | 45 |
| P8 | Tabuk | Male | Qabali Badawi | PhD | High | 50 |
| P9 | Riyadh | Male | Qabali Hadhri | PhD | High | 57 |
| P10 | Riyadh | Male | Qabali Hadhri | Bachelors | High | 60 |
| P11 | Makkah | Female | Arabi Hadhri | Masters | Mid | 38 |
| P12 | Makkah | Female | Arabi Hadhri | PhD | Mid | 39 |
| P13 | Makkah | Female | Arabi Hadhri | Masters | Low | 48 |
| P14 | Riyadh | Female | Qabali Hadhri | Some college | Low | 51 |
| P15a | Najran | Male | Qabali Badawi | Masters | Mid | 47 |
| P15b | Najran | Female | Qabali Badawi | Masters | Mid | 40 |

## 4.2 Inclusion Criteria

Even though we aimed to recruit Saudi Arabian parents that were at least 45 years or older and have a child that is at least 15 years old, 5 parents were younger than 45 and one of them had a child that was 14 years old. The reason we decided to be flexible with the inclusion criteria is to include more participants in our study. Also, this did not violate our rationale for these age thresholds, which is that parents would relate to their current role in finding a spouse for their young adult children rather than have them hypothesize future scenarios. Fifteen years is the age when a child would reach puberty from an Islamic perspective and are considered accountable adults [83]. They would not necessarily get married when they turn 15, but it would be the earliest age that the topic of marriage would be introduced. Also, this age is not associated with what parents perceived as an acceptable age for their children to use technology for marriage or any other purpose. Other inclusion criteria were that they are Saudi Arabian citizens or closely related to one (spouse or child) and have lived at least 10 years in Saudi Arabia.

## 4.3 Procedure

We conducted semi-structured interviews with 16 participants with 9 males and 7 females, which included one couple. Interviews were conducted either in person or over Skype, based on the participants' preference. Interviews were conducted in Arabic and lasted between 45 and 90 minutes and generated 87 pages of interview data. Participants were compensated with a gift that was equivalent to 50 Saudi Riyals or more. We were advised by a Saudi Arabian professional that it is more suitable to give gifts such as chocolate or fancy pens rather than cash to older participants. The interviews covered three aspects: participants initial thoughts about matchmaking technologies, reflecting on hypothetical features of matchmaking technologies, discussing the importance of sharing certain pieces of information through matchmaking technologies and concluding with any general comments or remarks they have about the study. Participants were provided a consent form both when signing up for the study as a soft copy and before conducting the interview as a hard copy. We stopped conducting interviews when data saturation was reached, and we did not gain new information from interviews.

### 4.4   Analysis

Interviews were transcribed in Arabic by the first and second authors. After that, the first author conducted qualitative inductive data driven thematic analysis. This began with open coding the interview transcriptions as mentioned in [40]. The interviews generated a total of 440 codes. After that, the first author used affinity diagraming [51] to generate thematic clusters by constantly comparing and organizing codes based on how similar they were  to each other. Thematic clusters were generated when a significant number of codes were highly similar to each other. During the process, clusters grew, shrunk, split and merged until the final set of clusters was generated and all codes were evaluated appropriately.

### 4.5 Research Position Statement

The first author was born in the US and raised between Saudi Arabia and the US and has dual citizenship. He spent his childhood and late adulthood in the US and his teenage years in Saudi Arabia. He is bilingual and is accustomed to most of the traditions of everyday life in Saudi Arabia. He is currently a PhD student at a US institution exploring ways to design digital technologies for Saudi Arabians to find their spouses online within cultural and religious expectations. His work is funded by a scholarship from Saudi Arabia.

## 5 RESULTS

Based on the thematic clusters resulting from our qualitative analysis, we identified three major themes which capture Saudi Arabian parents' perspectives on the potential use of matchmaking technologies. The first theme captures parental perspective regarding preserving their religious and cultural values and integrating them within technological systems. The second theme focuses on the parental belief that young adults are becoming more independent and autonomous when they use technology to seek a spouse for marriage. The third theme provides a rich accounting of parental concerns about how technology could be used or misused in the Saudi Arabian context.

### 5.1 Preserve Religious and Cultural Values Within Technological Systems

In this section, we report how Saudi Arabian parents sought to preserve their cultural and religious values when technological systems are involved in the marriage process. We begin with parents describing important religious and cultural aspects in the marriage process. After that, parents express how technology complements the traditional marriage process. Lastly, parents expect technology to preserve privacy from a cultural viewpoint.

*5.1.1 Managing Religious and Cultural Values.* Parents wanted to maintain the role that cultural and religious factors play in the process of choosing a suitable spouse for their young adults. For religion, being religious was preferred but it was more important that both spouses agreed on what it meant to be religious. For culture, it was important that families are culturally compatible to avoid conflicts and divorce. In addition, some cultural beliefs and practices negatively affect the traditional marriage process and technology might provide alternatives. Even though Saudi Arabians as a society might be skeptical about using technology for marriage [4], parents are hopeful that technology can play a positive role in the marriage process while respecting religious and cultural expectations.

*Religious Compatibility.* Religiosity is a source of reassurance for parents when they evaluate potential spouses for their young adults. They consider a potential spouse who is not religious to be more likely to cause injustice in a relationship. A mother, P15b, states "you know the person fears god" when describing a potential spouse who is religious. A potential spouse being religious is viewed as a sign of a successful marriage. It is typical in Saudi Arabian culture that potential spouses are asked about their religiosity as an initial filter to decide if they are a suitable candidate for marriage or not. A father, P9, explains that choosing a religious wife "might have a positive effect" on his son. This is supported by another father, P5, that prefers to marry off his son or daughter to a religious person even if they are not religious themselves. Parents viewed religion as a binary concept, with a religious person being "good" and a non-religious person being "bad." This is also associated with the marriage's success or failure.

*"Even if my son is not good and the girl is good, I prefer it. Better than a bad girl like him…or my daughter is not good, I marry her to a good man better than another because he is the best choice" P5*

On the other hand, religious compatibility between potential spouses was important to avoid conflicts in the future. A father, P6, emphasizes the importance of religious compatibility for his daughter's prospective proposals, "If an extremely liberal family proposed for my daughter, I would not accept." As a religious father, he would not want his daughter to feel obligated to assimilate with her potential husband's liberal family or vice versa. A mother, P13, argues that "If there is a religious conflict, it would be hard for [their marriage] to continue." This religious conflict arises because of different interpretations of what being religious means. Parents view a person's religiosity as an extension of their expectations and lifestyle, which may cause conflict when potential spouses have different expectations.

*"What type of hijab is required and other religious principles so they would help each other because religion is an important factor. Also, so they would not be surprised with something different in their lifestyle later." P14*

Overall, parents view a religious potential spouse as an indicator of a successful marriage but worry that conflicting interpretations of religiosity between spouses might lead to a tough marriage or divorce. Conflicting interpretations arise when cultural and religious values become interleaved. For example, in the previous quote, P14 mentions "What type of hijab is required," which has both a cultural and religious component. Some view *niqab* as religiously optional or a cultural obligation while others believe it is religiously required. As a result, both cultural and religious compatibility are needed.

*Culture Compatibility.* Parents emphasized that it was important that families are culturally compatible and were against certain cultural practices and beliefs that hindered the marriage process but did not want to stand out.

One of the most important cultural factors in Saudi Arabian marriages is tribal lineage. A mother who is Qabali, P4, explains "A man that has no tribal association, it is hard to accept him." The reason this matters for the parents is that they do not want their grandchildren to be outcasts because their father is Arabi. Parents generally prefer that spouses are either both Qabali or both Arabi to avoid problems that could affect their extended family. For example, Qabali families might boycott marrying from a certain Qabali family because one of their members married an Arabi. In some of these cases, the extended family might pressure the spouses or their parents to end the marriage to restore order within the family. In addition to that, identifying a person as Qabali or Arabi is not easy. Parents are reluctant to use technology for marriage because they are not sure if tribal lineage can be easily verified online as P2 states "Names are similar, like our [family name] can be tribal or not and you cannot tell." Compatible linage is crucial as it affects potential spouses and their extended family's future prospects.

> *"Compatible linage is very important because it affects the family. It affects my sisters, their daughters and female cousins." P2*

Another reason parents prefer the traditional marriage process is that it was easier to identify a Badawi or Hadhri lifestyle. The conflict of lifestyles could be a concern for the potential spouses. A mother, P2, recounts her experience "I married a Badawi and I am Hadhri and had a hard time. In the end I asked for *khul'*." In some cases, the potential spouse is suitable as an individual but the community they belong to has an undesirable gender hierarchy that is imposed by their lifestyle.

> *"When a great guy proposes to your daughter, but you know...their community demeans women. Why make my daughter deal with this? I do not want this environment" P6*

While it may seem discriminatory to not accept marriage between Qabalis and Arabis or Badawis and Hadhris, parents believed that the differences will eventually cause issues for their young adults and lead to separation or a bad marriage. From a religious standpoint, these cultural values conflict with Islamic teachings that are against discrimination. Nonetheless, they are still socially and culturally present in Saudi Arabia.

*Technological Alternatives are stigmatized.* While cultural practices such as spouses not meeting before the wedding night can lead to divorce, technological alternatives are stigmatized in the Saudi Arabian community. Parents are also concerned that the traditional method has led to many divorces. A mother, P4, explains "Each one blames the other and you know the reason is 'scratch and win'[4]." She believes divorce is a byproduct of spouses not knowing each other before the wedding night and compares it to buying a lottery ticket. Even though Saudi Arabian practices can create obstacles in the marriage process that parents do not necessarily agree with, they do not want to be the ones that break norms as it may have a negative effect on them and their young adults. This means parents realize that technology can be useful but are worried that other Saudi Arabians view it in a negative light. This in turn would threaten both the parents and their young adults' reputation.

> *"It is sensitive that someone knows that I allowed my daughter to use technology...It is like she is putting herself on display...there might be criticism from the community...I might allow it, but I will not let anyone know about it until it becomes common and [acceptable]" P12*

Parents realize that customs and culture have been changing recently and believe that technology might be able to help with the marriage process while preserving their values. With the changes in social dynamics over time, parents consider technology as a leverage they could utilize. A father, P10, states "Because of social circumstances, social connections have weakened, and people got busy...Technology might help in finding solutions for marriage seekers." Parents do not consider technology as a threat but rather that technology should be designed to improve the traditional marriage process while preserving and respecting their values. This becomes crucial when cultural and religious values conflict because of the different ways Saudi Arabians enact their religious and cultural values.

---

[4] 'Scratch and win' is a phrase used in Saudi Arabia to describe traditional marriages in a negative light.

*"Customs and culture have changed. Therefore, even in the process of [marriage], it is possible that [technology] serves us, such that we do not impinge on our religion, customs and culture." P3*

*Conclusion.* Parents were concerned about maintaining their cultural and religious values during the marriage process. Religious potential spouses were favored if they had compatible religious views. Having a similar linage and lifestyle was culturally preferred. Some cultural practices and beliefs created obstacles in the marriage process, but parents did not want to break the norms. Maintaining both cultural and religious values while avoiding conflicts between them is a complex process. These findings address some of the concerns and views Saudi Arabian parents have about marital matchmaking technologies (RQ1). Nonetheless, Parents are hopeful that technology can play a role in the matrimonial matchmaking process if it preserves and respects their values.

*5.1.2 Technology Enhances the Traditional Marriage Process.* Technology can improve the traditional marriage process, but the former should not cause the latter to become obsolete. Technology allows potential spouses to know each other better before marriage while the traditional marriage process maintains cultural norms and involves parents in the process.

*Access to Information.* The traditional marriage process limits potential spouses from exchanging information and representing themselves appropriately and technology could provide alternative solutions. Those who try to seek more information than what the traditional process provides are viewed negatively. Technology would allow potential spouses, especially women, to decide how they present themselves appropriately for marriage. A father, P8, explains "Women might need [technology] more…some present themselves in a bad way in weddings." The traditional marriage process utilizes weddings and social gatherings as a medium for women to present themselves to mothers of potential grooms. Technology is viewed as a better alternative for potential spouses to present themselves and exchange information compared to the traditional marriage process.

*"There is a lack of information when someone looks for a spouse. [Technology] might help with solving problems. Culturally, there are limitations and there is social pressure for those who try to seek information. Sometimes they seem like they have crossed boundaries or are attempting to seek people's secrets." P7*

*Technology to Traditional Transition.* Parents realize that young adults may utilize technology without their knowledge as it provides them with more options but insist that it must transition to the traditional process. Parents are against technology when it is used to replace the traditional process. A father, P8, says "There is no way [a dad] comes and proposes for his son and tells [me] they knew about my daughter through technology." Parents prefer to maintain the traditional process publicly even if technology was used privately. Parents are willing to help their young adults by convincing other parents to transition from technology to the traditional process. A father, P7, explains "I would meet her father and try to convince him that [they] would meet…in her house or a public place." Parents might accept that their young adults utilize technology for marriage without their knowledge if it transitions into the traditional process.

*"Technology would provide them with the comfort of freedom of choice. Some families are more accepting of technology. Sometimes the youth use it without telling their families." P12*

Parents consider their involvement in the transition from technology to be essential to make sure it aligns with cultural expectations. Parents viewed themselves as secondary screeners for potential spouses that are chosen by their young adults through technology so that societal standards are not affected. A father, P10, elaborates "I must sit with [potential grooms] to know everything...social status...intellectual compatibility and social compatibility." Parents usually want to be involved in the transition from technology to the traditional process to ensure it is done properly.

> *"In the beginning, it will be only the two knowing each other and understanding each other more but after that when it transitions into marriage, the families must intervene. They must educate each other about their family's norms [and traditions]" P12*

*Technology Utilized when Tradition is not Viable.* Parents prefer the traditional route when social connections are strong within the community but realize that technology can help widen the pool of potential spouses. A mother, P12, believes that technology might help spouses find each other within their own community that they would not consider otherwise. She says "[Technology] might be beneficial in making a person notice... [It] might say that this girl and her cousin are compatible." Parents would find technology helpful when they live in a community that they are not familiar with. A father, P10, says "[The father] lives in a community where no one knows him... if he went to technology, it would benefit him." Parents do consider technology to be helpful but prefer the traditional process as it is more efficient when it is feasible.

> *"A person living close to his family and community and everyone knows him...I think technology will not help. Traditional is better and faster." P10*

*Conclusion.* Parents prefer the traditional marriage process (without technology's facilitation), even though they realize that technology can help improve it. Parents major concern is that technology would replace the traditional process, which in turn would minimize their role as parents in the matrimonial matchmaking process. This addresses (RQ2) what role Saudi Arabian parents would play in matchmaking technologies. They emphasize that their role is to ensure that matchmaking technologies are only a tool to find a suitable potential spouse and that they lead to a transition to the traditional marriage process.

*5.1.3 Managing privacy online to protect young adults and preserve cultural values.* Parents are concerned that technology might expose young adults to dangerous situations or cause them to violate cultural expectations. Young adults should take precautions by not disclosing their identity online until they find a suitable spouse. At the same time, they should know enough about potential spouses to ensure they meet cultural standards. Daughters should be more vigilant about sharing their information online as it is more culturally sensitive compared to sons. In general, sharing excessive information can attract unintended harm to young adults. Parents want to protect their young adults and ensure that cultural expectations are preserved.

Parents believe that young adults have more control over the marriage process and can preserve their privacy online when they use technology anonymously. Some young adult's might be more sensitive about keeping their identity confidential and preserving their privacy online compared to others. A mother, P2, says "Some might die if their name is known...it is respectful to preserve someone's privacy." At the same time, technology enables young adults to get to know their potential spouse before they decide to get engaged to them traditionally and without their parents' knowledge. A mother, P12, says "You can choose that your mom and dad do not know because at first it is getting to know the person, not an engagement." Being anonymous also allows young adults to make multiple attempts to find a suitable spouse without worrying about their identity being revealed. A father, P10, states "A person used technology once, twice and thrice...when his identity is unknown, I think it would help this person to use it until he finds the suitable person." Being anonymous allows young adults to find a suitable spouse online and decided when to transition to the traditional marriage process while maintaining their privacy.

*"First talk about finances, age, school and tribal compatibility...in the beginning, a person would not want others to know him. After the family and community are compatible, we start an individual agreement." P4*

While being anonymous has its benefits, knowing a potential spouse's tribal affiliation and geographical location is essential for parents and young adults to decide if they are culturally compatible. Tribal affiliation should be known earlier in the process as it can make or break the relationship. A father, P8, explains "When they know each other, it is revealed that she was not tribal...if it was known earlier, there would not be a relationship." Geographical location could indicate a cultural mismatch as a father, P10, states "Some do not like to marry from different regions...they like from their environment, their region." It could also indicate a transportation burden for potential spouses as a father, P1, explains "The region is important to know the commute. Be sure she is from your region. Like taking her to her family." At the same time, young adults should not know each other's precise locations as it may cause them to get involved in culturally unacceptable interactions. It is important to know a potential spouse's tribal affiliation and regional location to measure cultural compatibility but revealing their precise location may lead to culturally inappropriate interactions.

*"Even if they were engaged, I prefer that they do not know their locations before marriage...they might meet when engaged and we find it unacceptable before marriage." P11*

Parents are more concerned about preserving their daughters' identity and modesty online compared to their sons. A father, P3, says "Maybe the man does not mind putting his picture and specific details, but a woman might be more reserved." It is common for Saudi Arabian women to manage their privacy online based on how their family would react to it. A father, P7, explains "Privacy is always an obstacle...What if [her family] saw [her] picture online? They will kill [her]. Social pressure is involved." How a woman presents herself online reflects on how society perceives her and accepts her as an acceptable potential spouse.

*"Most people are fearful to marry a girl having a picture, uncovered...she would be liberal or has a strong personality. Not suitable as a mother or to handle responsibility...if it was a pseudo name and her identity is hidden, society would accept it. It is unusual for a girl to be bold. Not usual for a girl to impose herself or show her identity frankly and transparently. It is hard, let us be realistic." P15b*

Parents are concerned about young adults sharing information about their location or identity that could threaten their safety. For example, if a young adult rejects a potential spouse that is not suitable for them, the potential spouse might retaliate. A mother, P13, says "Maybe if someone knew my daughter and she rejected him he would hurt her or try to chase her." Young adults sharing their location could result in potential spouses using it to stalk them. A father, P6, explains "Location could be a tool for the stalkers. I know they are going to this store, let me go there." As a result, parents encourage young adults to be cautious when they share their information and make sure it is done in a safe and private manner.

*"Even if you share, you can share it safely. Like do not put coordinates, put a large circle...this the applications main responsibility, to preserve the privacy and safety of the users" P7*

*Conclusion.* Parents want to ensure that young adults are using technology appropriately when seeking potential spouses to remain safe, preserve their privacy and maintain cultural expectations. While technology provides young adults with more agency in the marriage process, parents are worried about the risks that may arise. This section addresses some of the concerns that Saudi Arabian parents have about using marital matchmaking technologies (RQ1). They believe that marital matchmaking technologies can be utilized appropriately if young adults follow cultural and safety guidelines.

## 5.2 Young Adults Autonomy Requires Maturity

Parents believe that young adults should be able to decide who they will marry if they maintain a respectful relationship with their parents and are mature enough to understand what marriage entails. While parents demand respect and young adults demand autonomy, they both need a middle ground where both can voice their opinions and their communication can flourishes. Parents are usually concerned that their young adults are not prepared enough for marriage, especially when it comes to understanding the opposite gender. At the same time, gender interactions and expectations have drastically changed recently, which affects the marriage process for both parents and young adults. Parents want to be involved and respected in the marriage process and understand that they should grant their young adults more autonomy as they become more mature and aware of the changes in gender roles and expectations.

*5.2.1 Respecting Parents vs. Young Adults' Autonomy.* Parents want to balance between granting their young adults' autonomy and maintaining their respected role in society. Young adults choosing who they marry helps maintain a healthy relationship with their parents. Parents value their young adults' autonomy and privacy, which was unheard of in previous generations. Parents seek their young adults trust to keep them aware about their technological interactions. That given, parents expect young adults to respect their opinions when they disagree and hope to reach a mutual understanding when it comes to choosing a potential spouse. Parents could resort to utilizing their legal or social authority when their young adults are not cooperating with them. Finding a middle ground where young adults can make their own decisions and parents' opinions are respected is a constant struggle.

Young adults should be able to decide who they want to marry as it could make the marriage process easier and maintains a healthy relationship between parents and young adults. Parents might consider young adults finding their own spouse as relieving their parents from the burden of finding a suitable spouse for them. A father, P1, says *"I have children and give them total freedom...they went and searched and made it easier for me."* Parents believe that their intervention is limited to providing advice and it is up to young adults to decide to take it or leave it. At the same time, parents believe that young adults need their support and guidance as they explore alternative methods. A mother, P14, says *"a person might accept alternative methods for marriage and young adults need the family and their support and guidance."* Parents do not want to threaten their relationship with their young adults because of disagreement on who the young adults want to marry.

*"I cannot intervene except by simply advising but he is free. He would be a man. Fully and mentally aware of his choices...I am not ready to lose my daughter or her love because of this...her decision and she does what she wants. I respect it. She did not like the advice, she does what she wants." P15b*

Parents intervening too much in the marriage process undermines young adults' autonomy. Parents cannot claim that their young adults can choose who to marry if they need their approval in the process. A father, P7, explains *"If the son or daughter need approval, then it means they are not given the full authority...the problem of choosing a spouse is that the mother and father intervene."* Parents realize that times have changed and what was acceptable for them is not necessarily acceptable for their children. A mother, P15b, explains *"The usual abuse that is the pressure of [parents] on their [young adults] in deciding a suitable spouse for them. Our time, me and their dad, is different from their time."* Young adults should be granted autonomy by being able to preserve some privacy in their personal lives away from their parents to build some trust between them when it comes to marriage.

*"I do not know everything about them…I do not have any of my children on snapchat and do not want to know it because I tell them spying is not a way. Because the prophet says do not spy…Trust is needed." P2*

Parents realize that they must open communication channels between them and their young adults to be involved in the technological marriage process. If parents try to prevent their young adults from utilizing technology for marriage, they will find ways to do it without their knowledge. It is better for parents to support their young adults to remain in the loop. A father, P9, explains *"If they are doing it anyways, do not be an obstacle. At least they would inform you about it."* Parents also realize that restrictions that were given during their time no longer apply and that they need to adapt to the change to remain relevant. A father, P3, says *"In our time it was unacceptable to talk to your father in this manner…The change that is happening is not in your hands. You either get along with it and see the positive aspects or…"* Parents want their young adults to trust them and keep them informed about their technological marriage endeavors.

*"I try to build trust between me and them. So, they can tell me what they are doing. Know how they are using this technology. Are they using it appropriately? Maybe they went in a different or wrong direction. Building trust between me and my children is really important." P12*

Young adults are expected to do their part too by respecting their parents' opinions when they disagree. Parents understand that their young adults might leave them out when they know parents would not approve of their actions. A mother, P2, says *"She does not consult her mother because she does not like her mother's opinions…I will not tell you because you will say no."* Parents want to maintain a space for dialog where parents and young adults discuss and attempt to persuade each other without either of them imposing their opinion on the other. A mother, P12, explains *"Young adults would try to convince their family that this choice is the most suitable…there would be acceptance from both ends and room for dialog and not imposing opinions on each other."* Parents want to be respected by their young adults to reach a middle ground where both are satisfied.

*"I cannot prevent my boys and girls because they are old and make their own decisions, but the family have their respect in our society and the authority to reject or accept. I cannot tell my son not to do it but if I am not satisfied, he will reconsider" P13*

When parents' opinions were disregarded by young adults, parents may resort to using their authority to assert their opinions. This applies to daughters as fathers still have the legal authority to approve or disapprove of the marriage. This has not changed with the recent changes in male guardianship laws to one of the father's relief, P8, as he says, *"They still have not removed guardianship for marriage."* The parents realize that they cannot intervene like they used to in the past, but their legal authority for their daughters is their last resort.

*"In the past the family intervenes…today no, they cannot. It is his personal decision; he deals with it. No, the girl you can refuse, why? Because some things are in your hand. The boy no, his matters are in his hands." P5*

*Conclusion.* Young adults' autonomy is important when it comes to deciding who they marry, and parents believe their opinions can be valuable when making that decision. Parents worry that young adults would exclude them from the process when they decide who to marry. Parents want to grant their young adults' autonomy but also prevent them from making bad decisions. Parents also refer to how family dynamics have changed compared to the past where their own parents had more influence on them when it came to marriage. Nonetheless, parents expect their young adults to make decisions that do not go against their parents' wishes. This section addresses some of the concerns that Saudi Arabian parents have about using marital matchmaking technologies (RQ1). While it is up to young adults to decide who they marry, parents can provide them with guidance and advice to make the right decision.

*5.2.2 Young adults need to be educated about the opposite gender.* Parents are concerned that their young adults are not mature and prepared for marriage especially when it comes to understanding the opposite gender and sexual interactions. Parents are worried that their young adults cannot handle the responsibility that comes with marriage as their view of it is superficial. Another thing parent are worried about is that young adults need to have set and clear goals and make sure they do not conflict with their potential partner. Parents want to find a suitable way to educate their young adults about the opposite gender and sex. Parents related to their own experiences on how they were not educated properly about sex and how it affected their marriage. Parents wanted to ensure that their young adults have a positive experience with their marriage and avoid mistakes that parents have committed in their own marriages especially for a culturally sensitive topic like sex.

Parents are worried that their young adults do not understand what marriage entails and view it superficially, which leads to divorce or a negative experience. If a young adult cannot manage their own individual responsibilities, they are not prepared to get married. A mother, P2, explains *"Who cannot handle responsibility cannot be trusted even for traditional marriage."* Young adults must understand what their goals are before they decide to get married. This would help them understand themselves better and avoid divorce that could be caused by potential spouses' goals not being compatible. A father, P9, says *"It is better that everyone knows their goals...because first you would educate the person. Second, decrease divorce rates"* Parents do not believe that incompatible goals mean that potential spouses are not suitable for each other but rather that it would make the marriage tougher. A mother, P2, says *"When they have shared goals, they can accomplish them quickly...but if they are different, they might struggle a bit, take time."* Parents are generally concerned that young adults have a distorted view about marriage and do not realize that until it is too late.

*"Marriage is not brand names and travel. I know a woman who married and went to France and bought everything she wanted and when she came back, she asked for divorce and said he is not suitable. Because he refused to buy a Rolex watch worth $40k...she does not know that marriage is a responsibility" P4*

While parents find that sexual education is essential for a successful marriage, they are concerned about how it would be gained appropriately. Islam stipulates sexual guidelines in marriage and parents are concerned that some might not be aware of them as P3 says *"For the religious limits in sexual interactions, some might make a mistake."* and he considers sexual education through religion as appropriate *"if they are religiously educated, they would be modestly sexually educated."* In addition, sexual education is not mandated in Saudi Arabian schools and parents called for programs to fill the gap. Parents were against potential spouses discussing sexual matters with the opposite gender before marriage as P13 says *"Our religion, values and customs do not allow cross gender dialog about [sex]"* and P8 adds *"Maybe they will say what about [premarital sex]?"* Parents reflected on their previous marriage experiences and admit that they could have benefitted from being educated properly about sex before marriage. A lack of sexual education could lead to a horrible marriage experience or a divorce that could have been avoided.

*"What ruined traditional marriages is [the lack of sexual education]...Now I teach my daughter because it is essential for a successful marriage. It ruined some of our marriages because we were not educated." P2*

Even though it is important for potential spouses to understand each other's sexual needs, it was not appropriate for it to be to be the only priority. When the process starts with sex as the driving factor, *"it will lose the credibility of both being serious"* according to P10. At the same time, given how sex is a sensitive topic in a conservative country like Saudi Arabia, potential spouses would be reluctant to share details about it. P7 states *"I do not think anyone will give a real answer"* when it came to how sex would be addressed for potential spouse, even though he agrees that it is a valid and important concern for potential spouses. Sex should come after other more important goals that would be achieved through marriage such as creating a family.

> *"The goal is to create a family, not pleasure...sexual pleasure should be secondary, not primary. Tranquility is important. You need comfort more than pleasure." P5*

*Conclusion.* Being mature and properly educated about marital responsibilities and priorities can make or break a marriage. Parents consider their young adults as irresponsible and not well educated about what a marriage requires. Parents are worried that their young adults are not fully aware of their actions and would only realize their mistake when it is too late. Marital matchmaking technologies role should ensure that potential spouses are aware of what marriage entails before they decide on a suitable spouse (RQ2).

*5.2.3 Change of gender expectations, roles and interactions.* Previous generations had stricter gender segregation rules in public spaces and defined gender roles in marriage. This made interacting with the opposite gender before marriage unlikely especially when meeting online was not an option. Young adults nowadays are more likely to interact with the opposite gender before marriage in public spaces as many Saudi women are joining the workforce. It has changed how they approach marriage and decide who they marry as gender roles have drastically changed and can no longer be assumed. In addition, online platforms allowed them to freely express their opinions, thoughts, identities, and different perceptions to the opposite gender. The increase in gender interactions has given young adults more opportunities to choose who they marry independently, but the process became more nuance with gender roles becoming more fluid rather than agreed upon.

The change in gender interactions has made traditional marriages undesirable for young adults as it usually means their interaction with the opposite gender is limited and controlled by their parents. Parents have realized that young adults are not interested in traditional marriages as a mother, P12, finds it a sequence of them being *"more aware"* of alternative options. She points out that young adults have opportunities that were *"not available for previous generations"*, and it therefore *"became easier to choose through interactions between genders, which now became more prevalent."* This also shows that the case of gender interactions being "more prevalent" seems to justify it. This subtle sense of justification can be easily questioned, as parents tend to explain why genders mixing can be beneficial, even when they are against it. *"I am honestly against mixing and openness, but there is one advantage. That you get to know the other before anything happens"*, says P4, a mother. Her phrase *"before anything happens"* shows a case where parents do not accept relationships as experiences, but as projects that lead to marriage. And where if the relationship is not to last; it shall not start. P4 also highlights that when cross-gender interact occur, potential spouses can evaluate each other intellectually, *"when you accept the person it is because you knew them intellectually."* Even though gender interactions are limited to certain settings such as work and school, these settings create a perfect environment for potential spouses to evaluate each other. Technology that is utilized for school or work makes it easier for potential spouses to transition to marriage eventually. The prevalence of technology and cross-gender interactions allows and encourages young adults to seek alternative ways to find a suitable spouse that would not be available in a traditional marriage.

*"The next generation will marry through technology…They started to talk with each other for work or school…the mature guy and gal will think about marriage because it is the goal for a proper relationship." P2*

Even though gender expectations about modesty and genders mixing are changing in Saudi Arabia, parents believe there is no consensus on them and that they can vary from one family to another. For example, a mother, P4, believes that *"As long as she has not married, she must be modest."* At the same time, she finds it reasonable that *"When she has children and becomes a mother…as she grows, she breaks boundaries."* Becoming a mother seems to loosen the expectations that are enforced on women as they are considered more mature. Another example is a mother, P2, stating that *"Some do not like it if their brothers come and she goes inside. No, cover up and stay with us"* as some families find it acceptable that genders mix as long as modesty is maintained. On the other hand, she states that women covering is changing and that *"In the meantime, most girls do not cover…they must get into the details."* She highlights that it is important to discuss expectations to ensure that there is no conflict between potential spouses. With recent changes in gender dynamics, potential spouses need to discuss their expectations and validate their assumptions to avoid any conflict in their future marriage and how they raise their own children.

*"Dad's principle is that his daughters cover up in a certain way or they do not show their face. The [mother] is from a family that were raised on showing their face and consider it normal. Here is a conflict of principles. It will cause a disagreement in the marriage. It is better that it is clear from the beginning." P12*

The traditional role of husbands being breadwinners has become controversial with the increase of Saudi Arabian women joining the workforce. A mother, P12, states that *"[Women] gained a form of independence"* and adds that *"[they do] not need a husband"* as she compares it to traditional roles where wives were dependent on their husbands. At the same time, she believes that *"for the man, nothing has changed"* as the husband is still expected to be the breadwinner, even if their wife is employed. As a result, she concludes that *"the burden remains on the man in marriage."* A mother, P2, explains that it is important to discuss financial expectations when both spouses are employed "even if it might be hurtful" as she believes that *"it is normal to be transparent."* It is worth noting that from an Islamic perspective, husbands are expected to provide for their wives [39], even if the wife is financially well-off. However, some cultural practices and Islamic rulings provide room for negotiations between spouses where income might be shared [84]. As potential spouses' sharing their expenses and incomes is still controversial in the public sphere, it is important that they discuss and clarify their financial roles and expectations in marriage.

*"A guy might say I am responsible to spend even if she is employed and another says if you are employed, I will not spend on you…it is not right or wrong but him or her decide based on that. A woman might say…if I [am employed] my money is mine. Maybe it is hers, but another person might say no, that does not work." P7*

Saudi Arabian women's traditional role of being caretakers of the household is being challenged when they choose to work. As stated by a father, P3, some men *"[believe] that a woman does not leave the house"* but he believes this creates a conflict as when *"she is employed, it is impossible that [they] are compatible."* He also links it to the man attempting to protect his authority in the marriage by stating that *"some men do not want her to work…[they want] to be controlling and the guardian."* Even when women working does not affect men's authority in the marriage, fathers believe there is a clash between mothers working and raising their children. As a father, P1, says that it is *"better to clarify she wants to work or not."* He highlights the issue of childcare that would conflict with the mother working *"The issue when she works is where to put the children."* It is important to note that childcare for working women is a relatively new concept in Saudi Arabia, which means it is not always available or affordable. As a result, P1, argues that *"if he is satisfied that she works, the children could be delayed."* His statements demonstrate his perception of a separated role of motherhood and work for mothers, and that if a woman wants both, having children would be postponed. Parents believe that it is important to discuss gender roles as the assumed traditional roles are no longer salient in Saudi Arabia.

*Conclusion.* With Saudi Arabia's public sphere changing rapidly in the last couple of years, cross-gender interaction and gender roles and expectations have changed drastically. Parents are aware that these differences did not exist during their time and realize it is important for young adults to be transparent about them as it definitely would affect their future marriage. As a result, it is important for marital matchmaking technologies to take into account these changes and ensure that potential spouses are aware of them and how they may be perceived differently by different potential spouses (RQ2).

## 5.3 Concerns About Technology's Risks, Reliability and Regulation

There were many concerns that technology could be used inappropriately or abused by young adults or others. These concerns stem from not being familiar with marital matchmaking technologies, which causes parents to be fearful of them. In addition, information found online was questionable and usually unreliable for young adults to make informed decisions when looking for a suitable spouse online. As a result, regulating the use of technology was considered as a suitable method to make finding a spouse online more effective for young adults. These concerns made parents feel hesitant about their young adults utilizing technology to find a suitable spouse.

*5.3.1 Afraid of Potential Technological Harms.* Parents were fearful about their young adults utilizing technology to find a potential spouse online. Their fears were mainly around three aspects: not trusting strangers and the potential harm they may cause, worrying that their daughters are more susceptible to harms online, being uncertain about potential spouses' seriousness about marriage.

*"I do not think our society is fully aware how to use technology…that it could be beneficial, effective and safe…I think it needs time." P10*

Parents worried about young adults seeking potential spouses online since they are not familiar with it and could not fully trust strangers online. Parents were skeptical of this unfamiliar way to get married. A father, P1, says he *"Never heard of people who got married via technology"* and a mother, P11, rejects it by saying *"I did not believe in this thing, and do not support it."* Even though she has heard that some people found this method successful, she argues that *"it is not suitable for our community."* Another mother, P13, argues that her concerns stem from the fear of the unknown *"I cannot guarantee the manners of people who will communicate with my sons and daughters."* Her fear of the unknown is linked to the fear of the potential harm as she says, *"I do not know how it would affect our family's privacy if they get in touch with a stranger."* When parents are not sure how finding a potential spouse online will pan out or who is behind it, they are usually more fearful than they are hopeful as it is considered an unexplored territory for them and they want to become more familiar with it.

*"Even though I do not expect it…it might become a phenomenon and I would like to be aware of it… there might be something positive or good. We always focus on the negatives." P8*

The fear of the unknown is more sensitive for most parents when it comes to their daughters searching for a spouse online compared to their sons as they view it to be riskier. One mother, P12, argues that *"the girl might be deceived or embarrassed or someone might manipulate her with words"* and as a result *"she would need guidance and monitoring from the family to avoid mistakes or being a victim."* A father, P15a, supports this by saying *"[women] are emotional about their decisions"* but a mother, P15b, argues otherwise *"I do not agree with this. Boys are emotional and make wrong choices too."* These conflicting perceptions might be related to the changes in cross gender interactions that were almost non-existent during the time of the parents. A mother, P15b, says *"5 years ago, I used to think differently"* as she realizes that her perception has change over time. Nonetheless, parents are more cautious when their daughters utilize marital matchmaking technologies to interacting with potential spouses that are strangers.

Parents are worried that marital matchmaking are not utilized for marriage and lack the seriousness that is present in the traditional process. A father, P9, believes that *"some people are serious about marriage on these websites, but the other side might be doing it for fun."* A mother, P12, supports this by saying that *"[young adults] use it for fun but not for marriage."* A mother, P11, says she is willing to reconsider her position on marital matchmaking technologies *"If he proposes directly and showed himself from the beginning to us, I might change my mind."* This is supported by a father, P6, who draws the line by saying *"if he did not propose to your dad, eliminate him"* as he considers the proposal as serious step in the right direction. He continues by saying that *"Some bad people took advantage of [technology]. They used it to blackmail."* On the other hand, he also believes that *"not everyone who uses technological means is bad...there are respectful people."* Determining if a user of a marital matchmaking technology is serious about marriage is crucial and challenging for parents compared to the traditional process.

*Conclusion.* Parents are concerned that using technology to find a potential spouse is risky. This concern is mostly rooted in the uncertainty that comes from interacting with unknown potential spouses. While parents are willing to be more open minded, they still cannot fully trust technology. Compared to the traditional marriage process, these potential spouses' intentions are not clear for young adults and parents (RQ1).

*5.3.2 Concerns About Information Reliability.* Deception or lack of transparency and seriousness are concerns that arise when young adults seek a potential spouse online. The traditional method relies on meeting potential spouses in person to validate their information and seriousness. Since technology lacks that aspect, parents are reluctant to trust information that is provided by technology without meeting potential spouses in person. A father, P9, explains *"Sometimes they doubt the available information if they do not meet and know."* The concern is that potential spouses might attempt to present themselves in a positive light, which might involve deception and questions their credibility. *"There is no credibility"* as a mother, P4, states. She believes that *"the man always lies or makes himself perfect"* and her concerned is that *"she believes him."* Nonetheless, P4 realize and admits that *"the problem here is the people and not technology."* Deception might not always be intentional as information presented online is usually selective, especially in Saudi Arabia where not all opinions are respected equally. A father, P7 says *"Even if it was real, it would be partial...our community is not open to accept all opinions."* In some cases, the concern is that potential spouses are not serious or transparent enough when providing information to potential spouses online. This makes the process less effective when potential spouses have different levels of seriousness and transparency. A father, P10, explains *"Those using technology must be serious when providing their information and be transparent and precise for it to be effective."* Parents are concerned that their young adults are not cautious enough about the authenticity of information provided by potential spouses online and if potential spouses are actually serious about marriage online.

Misusing or abusing technology when seeking a spouse online is a concern as it may lead to young adults being blackmailed. Parents linked the misuse of technology to it being accepted by the community as a mother P2 states *"Misuse, we have to underline it. Also, community acceptance."* Since using technology for marriage is a new concept, parents are concerned it may be utilized for extortion. *"It might have greed or extortion"* as a father, P5 states. He believes that *"because it is new and those who use it...use it negatively"* but he is also hopeful that at some point *"when something becomes successful, people will follow."* Parent hope that there is a way to prevent blackmail and exploitation and believe it is reliant on protecting young adults' information appropriately. A father, P7 supports this by saying *"You need to do it in a way so that nobody can exploit or blackmail. You need to find a method to not allow a person to take people's information."* Parents are concerned that technology might be utilized by others to inflict harm on their young adults, which causes them to discourage their young adults from finding a potential spouse through technology until it is proven successful.

*Conclusion.* Parents are worried that finding a potential spouse online is lacks transparency and could cause harm for their young adults. Measures need to be taken to increase transparency and minimize technological misuse. Marital matchmaking technologies must incorporate these measures to be effective (RQ1).

### 5.3.3 Regulation of technology is needed for it to be effective. Regulating marital matchmaking technologies was suggested by parents to circumvent their concerns and worries about their young adults' safety and adherence to religious guidelines. This could be done by involving a mediator that would oversee the process and ensure that it is properly conducted. The mediator could be a private entity or a public entity such as the government. The government is viewed as an authority that would prevent any misuse and is trusted by Saudi Arabians. Other ways of regulation were related to how marital matchmaking technologies restricted users from certain actions that could lead to misuse. While regulation might seem restrictive and intrusive, it helps create a safe environment for young adults to seek potential spouses without worrying about being harmed or compromising their religious beliefs.

Having a trusted authority that regulates marital matchmaking technology is considered a safety net that would prevent unacceptable behavior that would occur between potential spouses. Unacceptable behavior might relate to general misuse of technology or violating cultural and religious expectations. A father, P6, believes that *"Not every source is trustworthy"* when questioning who is behind a marital matchmaking technology as he believes that a "[well known] sheikh or a government entity" is considered a trustworthy source. It is important that the those *"who run them are known"* according to a mother, P2. She believes it guarantees that they are not those *"who might leak information"* as marital matchmaking technologies *"[have] confidential info."* As a result, the government or a trusted entity implementing the marital matchmaking technology is *"more assuring for parents"* according to a mother, P12, as she thinks it ensures that *"there is no manipulation."* On the other hand, a father, P15a, is concerned that corruption within government entities could allow people to *"know loopholes in the system [and abuse them]."* At the same time, he believes that *"if it was a private entity? It would be more negative."* It is important that the entity that implements a marital matchmaking technology in Saudi Arabia is known and trusted, which is more likely to be a public entity rather than a private one.

Limiting interactions on a marital matchmaking system can prevent misuse and ensure that interactions are aligned with religious values. While interacting with multiple potential spouses at the same time is consider an advantage, it could be a distraction from finding a suitable spouse. A mother, P4, says *"after choosing a specific person, there would be no options until I am done with them."* She believes that a potential spouse should be evaluated appropriately before moving on to another potential spouse. In addition, this would minimize abuse as a father, P7, says *"You need a strategy to not allow a user to be a player."* He believes that limiting a user's actions would prevent them from abusing the system. Even if users are not abusing the system, they might get distracted by the available options. *"Why get distracted?"* says a mother, P15b, as she believes young adults should focus on one person at a time and not *"get to know 100 people at the same time."* A father, P8, links this to religion as "the prophet said do not propose on your brother's proposal." He considers potential spouses interacting with each other as a form of proposal between them and proposals should not be simultaneous. Marital matchmaking technologies can minimize misuse and preserve religious values by controlling what actions are available for users.

   *Conclusion.* Marital matchmaking technologies should be regulated by a public entity with limited interactions between users. This would make marital matchmaking technologies safer and more aligned with Saudi Arabian religious values (RQ2).

## 6 DISCUSSION

Our findings indicate that parents want their young adults to preserve religious and cultural values, keep parents involved when choosing who they want to marry and understand that parents are attempting to help young adults remain safe in the technological facilitated process. We explain how these finding can guide future work and the design of suitable marital matchmaking technologies in Muslim and Arab cultures.

### 6.1 Technology Should be Religiously and Culturally Inclusive

As technology is usually perceived to be disruptive, parents were concerned that religious and cultural values would be compromised when young adults utilized marital matchmaking technologies to find a suitable spouse. Technology does not need to be disruptive to be successful as a previous study [65] indicated that adhering to traditional cultural values has led to matrimonial matchmaking websites being  more successful than dating websites.

   Our finding suggest that parents believed that marital matchmaking technologies should coexist with the traditional process as they would complement each other. The traditional process could benefit from the ease of access to information in marital matchmaking technologies. At the same time, parents were worried that sharing too much information online could harm their young adults. Partial anonymity allows young adults to exchange religious and cultural information to determine compatibility. This ensures that they protect their privacy and reputation since no identifiable data is shared. This partial anonymity would gradually fade away as more private information is exchanged between potential spouses to determine if they are individually compatible and transition to the traditional process. The implementation of partial anonymity that ensures a smooth transition from online to offline needs further exploration to understand it better.

### 6.2 Involve Other Stakeholders in Matchmaking Technologies

Parents want to be involved in the marriage process and maintain a good relationship with their young adults. While parents believe young adults could benefit from their guidance and wisdom, they understand that it is important for their young adults to independently decide who they want to marry. Parents demand to be respected stems from the concept of *bir alwalidayn*, which does not require that that young adults lose their autonomy. Finding a balance where parents can be involved in a marital matchmaking system without threatening young adults' autonomy is complicated. As parents have mentioned different variations within Saudi Arabian culture, parents' involvement in marital matchmaking technologies is not uniformly unique.

Our findings indicate that parents consider involvement to be either collaborative or transactional. An example for collaborative involvement is where parents mentioned that they would keep in touch with their young adults throughout the process by being aware of their actions and providing them with guidance. For transactional involvement, parents would vet potential spouses found online and transition to the traditional process to make things official. Collaborative involvement is not clearly defined compared to transactional involvement that is an extension of the traditional process. As a result, collaborative involvement could be determined through discussions and negotiations between parents and young adults while transactional involvement would be defined based on certain phases in the matchmaking process. Implementing these types of involvement in a technological system may assume that young adults need to cooperate with their parents. An alternative would be a negotiation system where young adults and their parents have different access based on their agreed upon level of parental involvement. For example, young adults could propose potential spouses to their parents to get initial feedback or parents may suggest potential spouses for their young adults to consider.

Both types of involvement would require further investigation to understand how they would be implemented in a marital matchmaking technology in Saudi Arabia. Even though there are existing marital matchmaking technologies for Muslims that have considered parents' transactional involvement (i.e. Minder), there have not been any studies in the HCI community that have explored their effectiveness in Muslim and Arab cultures and if they can be improved. Previous studies [65] have explored forms of collaborative involvement for Indian matrimonial websites. To the best of our knowledge, there have not been many studies in the HCI community that explore how collaborative involvement can be affective in matrimonial websites for Muslim and Arab cultures. In general, we encourage the HCI community to investigate involving other stakeholders in matchmaking technologies as previous studies [49] have shown that matchmaking technologies benefit from having a "community." We also encourage the HCI community to conduct more inclusive investigations, as mentioned by Sharma et al. [65], with understudied populations such as Muslims and Arabs.

### 6.3 Regulation of Matchmaking Technologies Makes Them Safer

Given that marital matchmaking technologies are not widely used in Saudi Arabia, parents are concern about them being safe and reliable. Some of these concerns have been mentioned in many previous studies [26,36,63] and are not unique to the Saudi Arabian context. However, these concerns are perceived differently when viewed through cultural and religious lens.

Our participants have mentioned involving the government to make marital matchmaking technologies safer, which has been shown to be successful in previous studies [65]. Unfortunately, the Saudi Arabian government has not been involved in any marital matchmaking technologies so far. This opens an opportunity to explore alternatives on how matchmaking technologies can emulate the government's involvement. A previous study by Masden and Edwards [49] mentioned how a community for dating sites might serve as a form of regulation similar to how Wikipedia articles are regulated. It is also possible that other stakeholders (i.e. family or friends) can be part of the community that regulates matchmaking technologies. This might serve as a suitable alternative for involving the government in matchmaking technologies, even though it lacks the authority to prosecute users.

Our findings indicate that information reliability and validity is another concern for parents. One way to address it is by providing a verification mechanism by matchmaking technologies similar to how Indian matrimonial websites verify users through their phone numbers and official documents [65] or other online platforms such as AirBnB [85]. It is not clear if these mechanisms can be implemented in Saudi Arabia as having a copy of a Saudi national ID card is prohibited by the Saudi Arabian ministry of interior [86]. Even though using a Saudi national ID for verification might be problematic, verification through phone numbers is a suitable alternative since all phone numbers are required by law to be linked to a Saudi National ID[5] [87]. Further investigations are needed to explore if these verification methods are possible and effective.

### 6.4 HCI Feminism

Previous work has introduced 6 core qualities for feminist HCI and the pluralism quality is the most relevant to our study [20]. While individualistic cultures value being independent, collectivist cultures value collaboration. Parents expressed more concerns about their daughters being safe when using technology for marriage and fathers were more likely intervene in their daughters' decision-making process. While these might seem restrictive from a Western standpoint, it could be viewed as favorable for some Arab and Muslim women. Even when Saudi Arabian Women are against parents controlling who they marry, they still adhere to Islamic and religious guidelines. Feminist ideology and talking about "private matters" is considered socially unacceptable in Saudi Arabia. As a result, Saudi Arabian women have to construct their own version of Islamic Feminism [11,38,54] where they define what women's empowerment means in a patriarchal and Islamic society. For example, Saudi Arabian women might avoid going against their fathers in a case of *'adhl* as they view following their father's wishes falls under *bir alwalidayn*. It does not mean that they do not value being able to choose who to marry, but rather give priority to their Islamic obligation to respect their parents. In fact, some Saudi Arabian women have stated that *"Islam has freed women before the West comes with its conventions"* as a protest against the term "feminism" [38]. While Saudi Arabian women still face many challenges in Saudi Arabia, they prefer to tackle them while adhering to cultural and religious values as they consider this approach to be more socially acceptable.

### 6.5 Limitations

One limitation of our study was the use of Twitter, word-of-mouth. and snowball sampling to recruit. While we aimed to recruit a diverse set of participants, most participants were either acquainted with the authors through some channel or tech-savvy enough to have found out about this study via Twitter. We had limited access to participants who did not meet those characteristics.

Due to the snowball sampling, our participants were not representative of the Saudi Arabian population geographically. Even though we attempted to recruit participants from Eastern region, which is one of the three major regions in Saudi Arabia, we were not successful.

Perhaps due to the bias introduced by recruiting via Twitter, most of our participants were open and accepting of using technology for marriage in Saudi Arabia. We are certain that there

---

[5] Saudi Arabia's communication and information technology commission also provides a service to check phone numbers linked to a Saudi National ID to prevent it from being misused

are Saudi Arabian parents that are more strongly against using technology for marriage, but they did not take part in our study. Thus, our findings may be skewed towards Saudi Arabian parents that are in favor of using technology for marriage.

Finally, we realize that there are secondary stakeholders and topics that we were not able to include in our investigation given its current scope, such as siblings or extended family members and the topic of polygamy. These secondary stakeholders might also serve as mediators between young adults and their parents, especially when parents are against using technology for marriage. While polygamy is practiced in Saudi Arabia, only one father and one mother from our participants brought it up when relating to their own experiences.

## 7 CONCLUSION

Our study has explored how Saudi Arabian parents perceive marital matchmaking technologies and what role they want play in them. Our interviews with 16 Saudi Arabian parents revealed that it is important to preserve Saudi Arabian values within marital matchmaking technologies, find ways for technology to coexist with the traditional marriage process and minimize potential harms through regulation. Our findings provide insights to guide the HCI community to be more culturally sensitive and inclusive, which aims to understand and account for human values in the design process. We also provide implications on how to design matchmaking technologies in this context based on preserving values, involving parents and ensuring young adults' safety. We outlined different avenues for future investigations in this context and provide our solutions. With this work, we provide a contextually grounded research study to benefit the future work of value-sensitive design in the CSCW community.

## ACKNOWLEDGMENTS

We would like to thank all participants for their time and sharing their insights with us. A special thanks for all reviewers that helped improve our paper. Rana Alhanaya's help in translating consent forms and questions during IRB procedures was greatly appreciated.

## REFERENCES

1. Norah Abokhodair, Sofiane Abbar, Sarah Vieweg, and Yelena Mejova. 2016. Privacy and Twitter in Qatar: Traditional Values in the Digital World. In *Proceedings of the 8th ACM Conference on Web Science* (WebSci '16), 66–77. https://doi.org/10.1145/2908131.2908146
2. Norah Abokhodair, Adam Hodges, and Sarah Vieweg. 2017. Photo Sharing in the Arab Gulf: Expressing the Collective and Autonomous Selves. In *Proceedings of the 2017 ACM Conference on Computer Supported Cooperative Work and Social Computing* (CSCW '17), 696–711. https://doi.org/10.1145/2998181.2998338
3. Norah Abokhodair and Sarah Vieweg. 2016. Privacy & Social Media in the Context of the Arab Gulf. In *Proceedings of the 2016 ACM Conference on Designing Interactive Systems*, 672–683.
4. Adel Al-Dawood, Norah Abokhodair, Houda El mimouni, and Svetlana Yarosh. 2017. "Against Marrying a Stranger": Marital Matchmaking Technologies in Saudi Arabia. In *Proceedings of the 2017 Conference on Designing Interactive Systems* (DIS '17), 1013–1024. https://doi.org/10.1145/3064663.3064683
5. Kecia Ali. 2015. *Sexual Ethics and Islam: Feminist Reflections on Qur'an, Hadith and Jurisprudence*. Oneworld Publications.
6. Salwa Abdel Hameed Al-Khateeb. 1998. Women, Family and the Discovery of Oil in Saudi Arabia. *Marriage & Family Review* 27, 1–2: 167–189. https://doi.org/10.1300/J002v27n01_11
7. Hashem Abdullah Almakrami. 2015. Online Self- Disclosure Across Cultures: A Study of Facebook Use in Saudi Arabia and Australia. Queensland University of Technology.
8. Tahani Obaid M. Alruwaili. 2017. Self Identity and Community Through Social Media: The Experience of Saudi Female International College Students in the United States. University of Northern Colorado, United States -- Colorado.
9. Yeslam Al-Saggaf. 2011. Saudi females on Facebook: An ethnographic study. *International Journal of Emerging Technologies and Society* 9, 1: 1.

10. Yeslam Al-Saggaf. 2016. An Exploratory Study of Attitudes Towards Privacy in Social Media and the Threat of Blackmail: The Views of a Group of Saudi Women. *The Electronic Journal of Information Systems in Developing Countries* 75. https://doi.org/10.1002/j.1681-4835.2016.tb00549.x

11. Tamara Alsheikh, Jennifer A. Rode, and Siân E. Lindley. 2011. (Whose) Value-sensitive Design: A Study of Long- Distance Relationships in an Arabic Cultural Context. In *Proceedings of the ACM 2011 Conference on Computer Supported Cooperative Work* (CSCW '11), 75–84. https://doi.org/10.1145/1958824.1958836

12. Soraya Altorki. 1977. Family Organization and Women's Power in Urban Saudi Arabian Society. *Journal of Anthropological Research* 33, 3: 277–287. https://doi.org/10.1086/jar.33.3.3630009

13. Anne Aula and Sasha Lubomirsky. 2008. Blocked Sites and Offensive Videos: The Challenges of Teen Computer Use. In *CHI '08 Extended Abstracts on Human Factors in Computing Systems* (CHI EA '08), 2757–2762. https://doi.org/10.1145/1358628.1358757

14. Ayman Bajnaid and Yeslam Al-Saggaf. 2017. Impression Formation on Matrimonial Sites. In *Proceedings of the 29th Australian Conference on Computer-Human Interaction* (OZCHI '17), 77–86. https://doi.org/10.1145/3152771.3152780

15. Ayman Naji Bajnaid. 2016. A Study of Online Impression Formation, Mate Preferences and Courtship Scripts among Saudi Users of Matrimonial Websites. Department of Media and Communication.

16. Ayman Naji Bajnaid and Tariq Elyas. 2017. Exploring the Phenomena of Online Dating Platforms Versus Saudi Traditional Spouse Courtship in the 21st Century. *Digest of Middle East Studies* 26, 1: 74–96. https://doi.org/10.1111/dome.12104

17. Ayman Bajnaid, Giuseppe Alessandro Veltri, Tariq Elyas, and Mahmoud Maqableh. 2018. A Quantitative Survey of Online Impression Formation and Mate Preferences among Saudi Users of Matrimonial Websites. *Modern Applied Science* 12, 11: 121. https://doi.org/10.5539/mas.v12n11p121

18. Ayman Bajnaid, Giuseppe Alessandro Veltri, Tariq Elyas, and Ra'ed Masa'deh. 2019. Utilizing Matrimonial Web sites Among Saudi Users: An Empirical Study. *Digest of Middle East Studies* 28, 1: 164–193. https://doi.org/10.1111/dome.12158

19. Ayman Bajnaid, Giuseppe Alessandro Veltri, Ala'a Hamdi Gharibeh, Ala'a Hamdi Gharibeh, and Mahmoud Maqableh. 2018. Qualitative Interviews of Using Matrimonial Websites among Saudi Users. *Modern Applied Science* 12, 11: 223. https://doi.org/10.5539/mas.v12n11p223

20. Shaowen Bardzell. 2010. Feminist HCI: taking stock and outlining an agenda for design. 10.

21. Haifa Binsahl and Shanton Chang. 2012. International Saudi Female students in Australia and social networking sites: what are the motivations and barriers to communication. In *ISANA International Academy Association Conference*, 1–12. Retrieved September 19, 2016 from https://www.researchgate.net/profile/Haifa_Binsahl/publication/261100072_International_Saudi _Female_Students_in_Australia_and_Social_Networking_Sites_What_are_the_motivations_and_barri ers_to_communication/links/0f3175333a1d295df2000000.pdf

22. Jeremy Birnholtz, Irina Shklovski, Mark Handel, and Eran Toch. 2015. Let's Talk About Sex (Apps), CSCW. In *Proceedings of the 18th ACM Conference Companion on Computer Supported Cooperative Work & Social Computing* (CSCW'15 Companion), 283–288. https://doi.org/10.1145/2685553.2685557

23. Johanna Brewer, Joseph "Jofish" Kaye, Amanda Williams, and Susan Wyche. 2006. Sexual Interactions: Why We Should Talk About Sex in HCI. In *CHI '06 Extended Abstracts on Human Factors in Computing Systems* (CHI EA '06), 1695–1698. https://doi.org/10.1145/1125451.1125765

24. John T. Cacioppo, Stephanie Cacioppo, Gian C. Gonzaga, Elizabeth L. Ogburn, and Tyler J. VanderWeele. 2013. Marital satisfaction and break-ups differ across on-line and off-line meeting venues. *Proceedings of the National Academy of Sciences* 110, 25: 10135–10140. https://doi.org/10.1073/pnas.1222447110

25. Finn Christensen and Juergen Jung. 2010. Global Social Interactions with Sequential Binary Decisions: The Case of Marriage, Divorce, and Stigma. *The B.E. Journal of Theoretical Economics* 10, 1. https://doi.org/10.2202/1935-1704.1652

26. Camille Cobb and Tadayoshi Kohno. 2017. How Public Is My Private Life? Privacy in Online Dating. In *Proceedings of the 26th International Conference on World Wide Web* (WWW '17), 1231–1240. https://doi.org/10.1145/3038912.3052592

27. Peter C. Dodd. 1973. Family Honor and the Forces of Change in Arab Society. *International Journal of Middle East Studies* 4, 1: 40–54.

28. Mohsen A F El-Hazmi, A R Al-Swailem, A S Warsy, A M Al-Swailem, R Sulaimani, and A A Al-Meshari. Consanguinity among the Saudi Arabian population. 5.

29. Andrew T. Fiore and Judith S. Donath. 2004. Online Personals: An Overview. In *CHI '04 Extended Abstracts on Human Factors in Computing Systems* (CHI EA '04), 1395–1398. https://doi.org/10.1145/985921.986073

30. Arup Kumar Ghosh, Karla Badillo-Urquiola, Mary Beth Rosson, Heng Xu, John M. Carroll, and Pamela J. Wisniewski. 2018. A Matter of Control or Safety?: Examining Parental Use of Technical Monitoring Apps on Teens' Mobile Devices. In *Proceedings of the 2018 CHI Conference on Human Factors in Computing Systems* (CHI '18), 194:1–194:14. https://doi.org/10.1145/3173574.3173768

31. Arup Kumar Ghosh, Karla Badillo-Urquiola, and Pamela Wisniewski. 2018. Examining the Effects of Parenting Styles on Offline and Online Adolescent Peer Problems. In *Proceedings of the 2018 ACM Conference on Supporting Groupwork* (GROUP '18), 150–153. https://doi.org/10.1145/3148330.3154519

32. Rosanna E. Guadagno, Bradley M. Okdie, and Sara A. Kruse. 2012. Dating deception: Gender, online dating, and exaggerated self-presentation. *Computers in Human Behavior* 28, 2: 642–647. https://doi.org/10.1016/j.chb.2011.11.010

33. Hala Guta and Magdalena Karolak. 2015. Veiling and blogging: social media as sites of identity negotiation and expression among Saudi women. *Journal of International Women's Studies* 16, 2: 115.

34. Jeffrey A. Hall. 2014. First comes social networking, then comes marriage? Characteristics of Americans married 2005-2012 who met through social networking sites. *Cyberpsychology, Behavior and Social Networking* 17, 5: 322–326. https://doi.org/10.1089/cyber.2013.0408

35. Jeffrey T. Hancock, Catalina Toma, and Nicole Ellison. 2007. The Truth About Lying in Online Dating Profiles. In *Proceedings of the SIGCHI Conference on Human Factors in Computing Systems* (CHI '07), 449–452. https://doi.org/10.1145/1240624.1240697

36. Mark J. Handel and Irina Shklovski. 2012. Disclosure, Ambiguity and Risk Reduction in Real-time Dating Sites. In *Proceedings of the 17th ACM International Conference on Supporting Group Work* (GROUP '12), 175–178. https://doi.org/10.1145/2389176.2389203

37. ADE HIDAYAT. THE VALUE INHERITANCE OF FAMILY SYSTEMS IN ISLAMIC TRADITION: 8.

38. Jacquelyn Lauren Hoza. 2019. Is There Feminism in Saudi Arabia? *UF Journal of Undergraduate Research* 20, 2. https://doi.org/10.32473/ufjur.v20i2.106192

39. Samia Ibtasam, Lubna Razaq, Maryam Ayub, Jennifer R. Webster, Syed Ishtiaque Ahmed, and Richard Anderson. 2019. "My cousin bought the phone for me. I never go to mobile shops.": The Role of Family in Women's Technological Inclusion in Islamic Culture. *Proceedings of the ACM on Human-Computer Interaction* 3, CSCW: 46:1–46:33. https://doi.org/10.1145/3359148

40. Irving Seidman. 1998. *Interviewing as Qualitative Research: A Guide for Researchers in Education And the Social Sciences.* Teachers College Press, New York.

41. Mikkel S. Jørgensen, Frederik K. Nissen, Jeni Paay, Jesper Kjeldskov, and Mikael B. Skov. 2016. Monitoring Children's Physical Activity and Sleep: A Study of Surveillance and Information Disclosure. In *Proceedings of the 28th Australian Conference on Computer-Human Interaction* (OzCHI '16), 50–58. https://doi.org/10.1145/3010915.3010936

42. Varda Konstam, Samantha Karwin, Teyana Curran, Meaghan Lyons, and Selda Celen-Demirtas. 2016. Stigma and Divorce: A Relevant Lens for Emerging and Young Adult Women? *Journal of Divorce & Remarriage* 57, 3: 173–194. https://doi.org/10.1080/10502556.2016.1150149

43. A. A. R. Kritem, A. Abou Rakba, and I. F. Al-Aissawi. 1981. Family Saudi: Role and changes and their impact on decision making (Arabic text). *King Abdul Aziz University, College of economics, Research Center.*

44. Adeline Y. Lee and Amy S. Bruckman. 2007. Judging You by the Company You Keep: Dating on Social Networking Sites. In *Proceedings of the 2007 International ACM Conference on Supporting Group Work* (GROUP '07), 371–378. https://doi.org/10.1145/1316624.1316680

45. Xiao Ma, Nazanin Andalibi, Louise Barkhuus, and Mor Naaman. 2017. "People Are Either Too Fake or Too Real": Opportunities and Challenges in Tie-Based Anonymity. In *Proceedings of the 2017 CHI Conference on Human Factors in Computing Systems* (CHI '17), 1781–1793. https://doi.org/10.1145/3025453.3025956

46. Xiao Ma, Jeff Hancock, and Mor Naaman. 2016. Anonymity, Intimacy and Self-Disclosure in Social Media. In *Proceedings of the 2016 CHI Conference on Human Factors in Computing Systems* (CHI '16), 3857–3869. https://doi.org/10.1145/2858036.2858414

47. Xiao Ma, Emily Sun, and Mor Naaman. 2017. What Happens in Happn: The Warranting Powers of Location History in Online Dating. In *Proceedings of the 2017 ACM Conference on Computer Supported Cooperative Work and Social Computing* (CSCW '17), 41–50. https://doi.org/10.1145/2998181.2998241

48. Christopher M. Mascaro, Rachel M. Magee, and Sean P. Goggins. 2012. Not Just a Wink and Smile: An Analysis of User-defined Success in Online Dating. In *Proceedings of the 2012 iConference* (iConference '12), 200–206. https://doi.org/10.1145/2132176.2132202

49. Christina Masden and W. Keith Edwards. 2015. Understanding the Role of Community in Online Dating. In *Proceedings of the 33rd Annual ACM Conference on Human Factors in Computing Systems* (CHI '15), 535–544. https://doi.org/10.1145/2702123.2702417

50. Laura Menin. 2020. 'Destiny is written by God': Islamic predestination, responsibility, and transcendence in Central Morocco. *Journal of the Royal Anthropological Institute* 26, 3: 515–532. https://doi.org/10.1111/1467-9655.13312

51. Michael Muller, Judith S. Olson, and Wendy A. Kellogg. 2014. Curiosity, Creativity, and Surprise as Analytic Tools: Grounded Theory Method. In *Ways of Knowing in HCI*. Springer New York, 25–48.

52. Diego Muñoz, Bernd Ploderer, and Margot Brereton. 2018. Towards Design for Renegotiating the Parent-adult Child Relationship After Children Leave Home. In *Proceedings of the 30th Australian Conference on Computer-Human Interaction* (OzCHI '18), 303–313. https://doi.org/10.1145/3292147.3292149

53. Diego Muñoz, Bernd Ploderer, and Margot Brereton. 2019. Position Exchange Workshops: A Method to Design for Each Other in Families. In *Proceedings of the 2019 CHI Conference on Human Factors in Computing Systems* (CHI '19), 109:1–109:14. https://doi.org/10.1145/3290605.3300339

54. Maryam Mustafa, Shaimaa Lazem, Ebtisam Alabdulqader, Kentaro Toyama, Sharifa Sultana, Samia Ibtasam, Richard Anderson, and Syed Ishtiaque Ahmed. 2020. IslamicHCI: Designing with and within Muslim Populations. In *Extended Abstracts of the 2020 CHI Conference on Human Factors in Computing Systems*, 1–8. https://doi.org/10.1145/3334480.3375151

55. Soud Nassir, Adel Al-Dawood, Elham Alghamdi, and Eman Alyami. 2019. "My Guardian Did Not Approve!": Stories from Fieldwork in Saudi Arabia. *Interactions* 26, 3: 44–49. https://doi.org/10.1145/3318145

56. R. Nayak, M. Zhang, and L. Chen. 2010. A Social Matching System for an Online Dating Network: A Preliminary Study. In *2010 IEEE International Conference on Data Mining Workshops*, 352–357. https://doi.org/10.1109/ICDMW.2010.36

57. Borke Obada-Obieh and Anil Somayaji. 2017. Can I Believe You?: Establishing Trust in Computer Mediated Introductions. In *Proceedings of the 2017 New Security Paradigms Workshop* (NSPW 2017), 94–106. https://doi.org/10.1145/3171533.3171544

58. Josue Ortega and Philipp Hergovich. 2017. The Strength of Absent Ties: Social Integration via Online Dating. *SSRN Electronic Journal*. https://doi.org/10.2139/ssrn.3044766

59. Anthony T. Pinter, Pamela J. Wisniewski, Heng Xu, Mary Beth Rosson, and Jack M. Caroll. 2017. Adolescent Online Safety: Moving Beyond Formative Evaluations to Designing Solutions for the Future. In *Proceedings of the 2017 Conference on Interaction Design and Children* (IDC '17), 352–357. https://doi.org/10.1145/3078072.3079722

60. John R. Porter, Kiley Sobel, Sarah E. Fox, Cynthia L. Bennett, and Julie A. Kientz. 2017. Filtered Out: Disability Disclosure Practices in Online Dating Communities. *Proc. ACM Hum.-Comput. Interact.* 1, CSCW: 87:1–87:13. https://doi.org/10.1145/3134722

61. POV. 2013. Infographic: A History of Love & Technology | xoxosms | POV | PBS. *POV | American Documentary Inc.* Retrieved September 17, 2019 from http://archive.pov.org/xoxosms/infographic-technology-dating

62. Mennatallah Saleh, Mohamed Khamis, and Christian Sturm. 2019. What About My Privacy, Habibi?: Understanding Privacy Concerns and Perceptions of Users from Different Socioeconomic Groups in the Arab World. In *Human-Computer Interaction – INTERACT 2019* (Lecture Notes in Computer Science), 67–87.

63. Nithya Sambasivan, Amna Batool, Nova Ahmed, Tara Matthews, Kurt Thomas, Laura Sanely Gaytán-Lugo, David Nemer, Elie Bursztein, Elizabeth Churchill, and Sunny Consolvo. 2019. "They Don't Leave Us Alone Anywhere We Go": Gender and Digital Abuse in South Asia. In *Proceedings of the 2019 CHI Conference on Human Factors in Computing Systems* (CHI '19), 1–14. https://doi.org/10.1145/3290605.3300232

64. Diane J. Schiano, Christine Burg, Anthony Nalan Smith, and Florencia Moore. 2016. Parenting Digital Youth: How Now? In *Proceedings of the 2016 CHI Conference Extended Abstracts on Human Factors in Computing Systems* (CHI EA '16), 3181–3189. https://doi.org/10.1145/2851581.2892481

65. Vishal Sharma, Bonnie Nardi, Juliet Norton, and A. M. Tsaasan. 2019. Towards Safe Spaces Online: A Study of Indian Matrimonial Websites. In *Human-Computer Interaction – INTERACT 2019* (Lecture Notes in Computer Science), 43–66. https://doi.org/10.1007/978-3-030-29387-1_4

66. Ramina Sotoudeh, Roger Friedland, and Janet Afary. 2017. Digital romance: the sources of online love in the Muslim world. *Media, Culture & Society* 39, 3: 429–439. https://doi.org/10.1177/0163443717691226

67. C. L. Toma, J. T. Hancock, and N. B. Ellison. 2008. Separating Fact From Fiction: An Examination of Deceptive Self-Presentation in Online Dating Profiles. *Personality and Social Psychology Bulletin* 34, 8: 1023–1036. https://doi.org/10.1177/0146167208318067

68. Ralph Vacca. 2019. Brokering Data: Co-Designing Technology with Latina Teens to Support Communication with Parents: Leveraging Cultural Practices of Latinx Youth Through Co-Design. In *Proceedings of the 18th ACM International Conference on Interaction Design and Children* (IDC '19), 197–207. https://doi.org/10.1145/3311927.3323142

69. As Warsy, Mh Al-Jaser, A Albdass, S Al-Daihan, and M Alanazi. 2014. Is consanguinity prevalence decreasing in Saudis?: A study in two generations. *African Health Sciences* 14, 2: 314. https://doi.org/10.4314/ahs.v14i2.5

70. Pamela Wisniewski, Arup Kumar Ghosh, Heng Xu, Mary Beth Rosson, and John M. Carroll. 2017. Parental Control vs. Teen Self-Regulation: Is There a Middle Ground for Mobile Online Safety? In *Proceedings of the 2017 ACM Conference on Computer Supported Cooperative Work and Social Computing* (CSCW '17), 51–69. https://doi.org/10.1145/2998181.2998352

71. Pamela J. Wisniewski, Heng Xu, Mary Beth Rosson, and John M. Carroll. 2014. Adolescent Online Safety: The "Moral" of the Story. In *Proceedings of the 17th ACM Conference on Computer Supported Cooperative Work & Social Computing* (CSCW '14), 1258–1271. https://doi.org/10.1145/2531602.2531696

72. Pamela Wisniewski, Haiyan Jia, Na Wang, Saijing Zheng, Heng Xu, Mary Beth Rosson, and John M. Carroll. 2015. Resilience Mitigates the Negative Effects of Adolescent Internet Addiction and Online Risk Exposure. In *Proceedings of the 33rd Annual ACM Conference on Human Factors in Computing Systems* (CHI '15), 4029–4038. https://doi.org/10.1145/2702123.2702240

73. Pamela Wisniewski, Heng Xu, Mary Beth Rosson, and John M. Carroll. 2017. Parents Just Don'T Understand: Why Teens Don'T Talk to Parents About Their Online Risk Experiences. In *Proceedings of the 2017 ACM Conference on Computer Supported Cooperative Work and Social Computing* (CSCW '17), 523–540. https://doi.org/10.1145/2998181.2998236

74. Doug Zytko, Sukeshini Grandhi, and Quentin (Gad) Jones. 2016. The Coaches Said...What?: Analysis of Online Dating Strategies Recommended by Dating Coaches. In *Proceedings of the 19th International Conference on Supporting Group Work* (GROUP '16), 385–397. https://doi.org/10.1145/2957276.2957287

75. Doug Zytko, Jessa Lingel, Jeremy Birnholtz, Nicole B. Ellison, and Jeff Hancock. 2015. Online Dating As Pandora's Box: Methodological Issues for the CSCW Community. In *Proceedings of the 18th ACM Conference Companion on Computer Supported Cooperative Work & Social Computing* (CSCW'15 Companion), 131–134. https://doi.org/10.1145/2685553.2699335

76. Douglas Zytko. 2016. Enhancing Evaluation of Potential Romantic Partners Online. In *Proceedings of the 19th International Conference on Supporting Group Work* (GROUP '16), 517–520. https://doi.org/10.1145/2957276.2997030

77. Douglas Zytko, Sukeshini A. Grandhi, and Quentin Jones. 2016. Online Dating Coaches' User Evaluation Strategies. In *Proceedings of the 2016 CHI Conference Extended Abstracts on Human Factors in Computing Systems* (CHI EA '16), 1337–1343. https://doi.org/10.1145/2851581.2892482

78. Douglas Zytko, Victor Regalado, Sukeshini A. Grandhi, and Quentin Jones. 2018. Supporting Online Dating Decisions with a Prompted Discussion Interface. In *Companion of the 2018 ACM Conference on Computer Supported Cooperative Work and Social Computing* (CSCW '18), 353–356. https://doi.org/10.1145/3272973.3274095

79. 2019. Saudi Arabia ends gender segregation in restaurants. *BBC News*. Retrieved September 10, 2020 from https://www.bbc.com/news/world-middle-east-50708384

80. Historic day as Saudi women get behind the wheel to drive | Al Arabiya English. Retrieved September 10, 2020 from https://english.alarabiya.net/en/News/gulf/2018/06/24/-SaudiWomenDriving-officially-comes-into-effect-across-the-kingdom

81. Revamped guardianship laws usher in a new era for Saudi women | Al Arabiya English. Retrieved September 10, 2020 from https://english.alarabiya.net/en/features/2019/08/03/Revamped-guardianship-laws-usher-in-a-new-era-for-Saudi-women

82. Saudi Arabia: social network penetration 2017 | Statistic. *Statista*. Retrieved March 1, 2018 from https://www.statista.com/statistics/284451/saudi-arabia-social-network-penetration/

83. If any of the signs of puberty appear in a boy, he becomes accountable - Islam Question & Answer. Retrieved December 8, 2019 from https://islamqa.info/en/answers/197392/if-any-of-the-signs-of-puberty-appear-in-a-boy-he-becomes-accountable

84. Is the husband obliged to spend on his wife if she is working? Does he have the right to take anything of her salary? - Islam Question & Answer. Retrieved September 2, 2020 from https://islamqa.info/en/answers/126316/is-the-husband-obliged-to-spend-on-his-wife-if-she-is-working-does-he-have-the-right-to-take-anything-of-her-salary

85. ID and verification | Airbnb Help Center. *Airbnb*. Retrieved May 30, 2020 from https://www.airbnb.com/help/topic/1389/id-and-verification

86. Civil Affairs. Retrieved May 30, 2020 from https://www.moi.gov.sa/wps/portal/Home/sectors/civilaffairs/contents/!ut/p/z1/jZHNUoMwFEZfhQ1LyS2goLsMMwK1jqJDxWwYKCnEgaRNg9i3N8VudCxtVrnJ-fJzLiIoQ4QXn6wuFBO8aHX9Tm5yiF03mrn2g-8-3wNO7HmAg9gOHz30NgJBiCPXWwD4i_AaYhylL7eJ4wB2ELkkDycGhgvzpwEyffxrIdH83CXaAvvYbglGZCW4ol8KZZ1gRsWGgptwmDaio8a4yZUJh8oESeu-HU2OTM6PVg1W5RVdF32rcsr1F8jkA7D7FwDsawA7yyjyEvvJnx2BqTacE6El1K0of3qOeen4NSKSrqmk0uqlXm6U2uzuTDBhGAarFqJuqbUSnQn_RRqx05Z-k2jTpWk27B1g8RUp98M3msHkSg!!/dz/d5/L0lHSkovd0RNQUprQUVnQSEhLzROVkUvZW4!/

87. CITC E-Services. Retrieved May 30, 2020 from https://portalservices.citc.gov.sa/E-Services/MyNumbers/MyNumbersInquiry.aspx